\documentclass[aps,pre,twocolumn,showpacs,floatfix,groupedaddress]{revtex4}

\usepackage{algorithm}
\usepackage{algpseudocode}
\usepackage{amsmath}
\usepackage{amsthm}
\usepackage{amssymb}
\usepackage{amscd}
\usepackage{bm}
\usepackage{latexsym}
\usepackage{mathtools}
\usepackage{comment}

\usepackage{graphicx}
\usepackage{epsfig}
\usepackage{epstopdf}
\usepackage{svg}
\usepackage[caption=false]{subfig}
\usepackage{placeins}

\renewcommand{\vec}[1]{\bar{#1}}
\providecommand{\mat}[1]{\bar{\bar{#1}}}

\providecommand{\config}[1]{\mathbf{#1}}

\providecommand{\abs}[1]{\vert#1\vert}

\providecommand{\RR}{\mathbb{R}}
\providecommand{\config}{{\bf C}}
\providecommand{\eqref}[1]{(\ref{#1})}

\begin{document}
	
	\title{Quotient Maps and Configuration Spaces of Hard Disks}
	
	\author{O. B. Eri\c{c}ok}
	\email{oericok@ucdavis.edu}
	\affiliation{Materials Science and Engineering, University of California, Davis, CA, 95616, USA.}
	
	\author{J. K. Mason}
	\email{jkmason@ucdavis.edu}
	\affiliation{Materials Science and Engineering, University of California, Davis, CA, 95616, USA.}
	
	\begin{abstract}
		Hard disks systems are often considered as prototypes for simple fluids. In a statistical mechanics context, the hard disk configuration space is generally quotiented by the action of various symmetry groups. The changes in the topological and geometric properties of the configuration spaces effected by such quotient maps are studied for small numbers of disks on a square and hexagonal torus. A metric is defined on the configuration space and the various quotient spaces that respects the desired symmetries. This is used to construct explicit triangulations of the configuration spaces as $\alpha$-complexes. Critical points in a configuration space are associated with changes in the topology as a function of disk radius, are conjectured to be related to the configurational entropy of glassy systems, and could reveal the origins of phase transitions in other systems. The number and topological and geometric properties of the critical points are found to depend on the symmetries by which the configuration space is quotiented.
	\end{abstract}
	
	
	\maketitle

	\section{Introduction}
	\label{sec:introduction}

	The glass transition is a subject of ongoing study in condensed matter physics. Since it is related to a slowing down of the dynamics and is not accompanied by a change in any obvious structural order parameter, it is usually not considered to be a true thermodynamic phase transition. Recent computer simulations \cite{berthier2019perspective} suggest that the main difference between a glass and a liquid is the volume of configuration space that is available to both systems. The volume of configuration space relevant to a glassy system is often supposed to be proportional to the number of local minima of the potential energy surface. An accurate count of these minima would then allow the configurational entropy to be used as an order parameter \cite{berthier2019perspective}, and a popular strategy to enumerate potential energy minima was proposed by Goldstein \cite{goldstein1969viscous} and formalized by Stillinger and Weber \cite{stillinger1982hidden,stillinger1995topographic}. The assumption underlying this view of the glass transition is that each local minimum of the potential energy surface corresponds to a different glassy state.
	
	Local minima are specific examples of a larger class of points known as critical points, roughly defined as locations where the topology of a manifold changes. The number and distribution of critical points of the potential energy surface has also been implicated in the onset of phase transitions, an idea known as the Topological Hypothesis \cite{caiani1997geometry,franzosi2004theorem}. Consider a system of particles with positions $\vec{q}_i$ and potential energy $V(\vec{q}_1,\dots,\vec{q}_N)$. Previously, Franzosi et al.\ \cite{franzosi2007topology1,franzosi2007topology2,gori2018topological} claimed that a change in the topology of the equipotential energy submanifolds $\Sigma_\nu = V^{-1}(-\infty, \nu]$ of the configuration space as a function of the energy $\nu$ is a necessary condition for a phase transition to occur in systems with smooth, stable, confining, and short-range interactions. Kastner and Mehta \cite{kastner2011phase} eventually found a counterexample satisfying all the stated conditions, but for which a phase transition occurs without a change in topology. They then proposed new criteria stating that a phase transition requires either (i) the number of critical points in a narrow potential energy band to grow exponentially faster than the number of particles, or (ii) the determinant of the Hessian matrix to vanish for a significant fraction of the critical points. It is significant that the Topological Hypothesis, either the original or the revised one, has so far only been evaluated for systems simple enough to be treated at least partially analytically; there appears to not yet even be the machinery available to test the hypothesis for, e.g., a simple fluid.
	
	\begin{figure}[b]
		\centering
		\includegraphics[width=1.0\columnwidth]{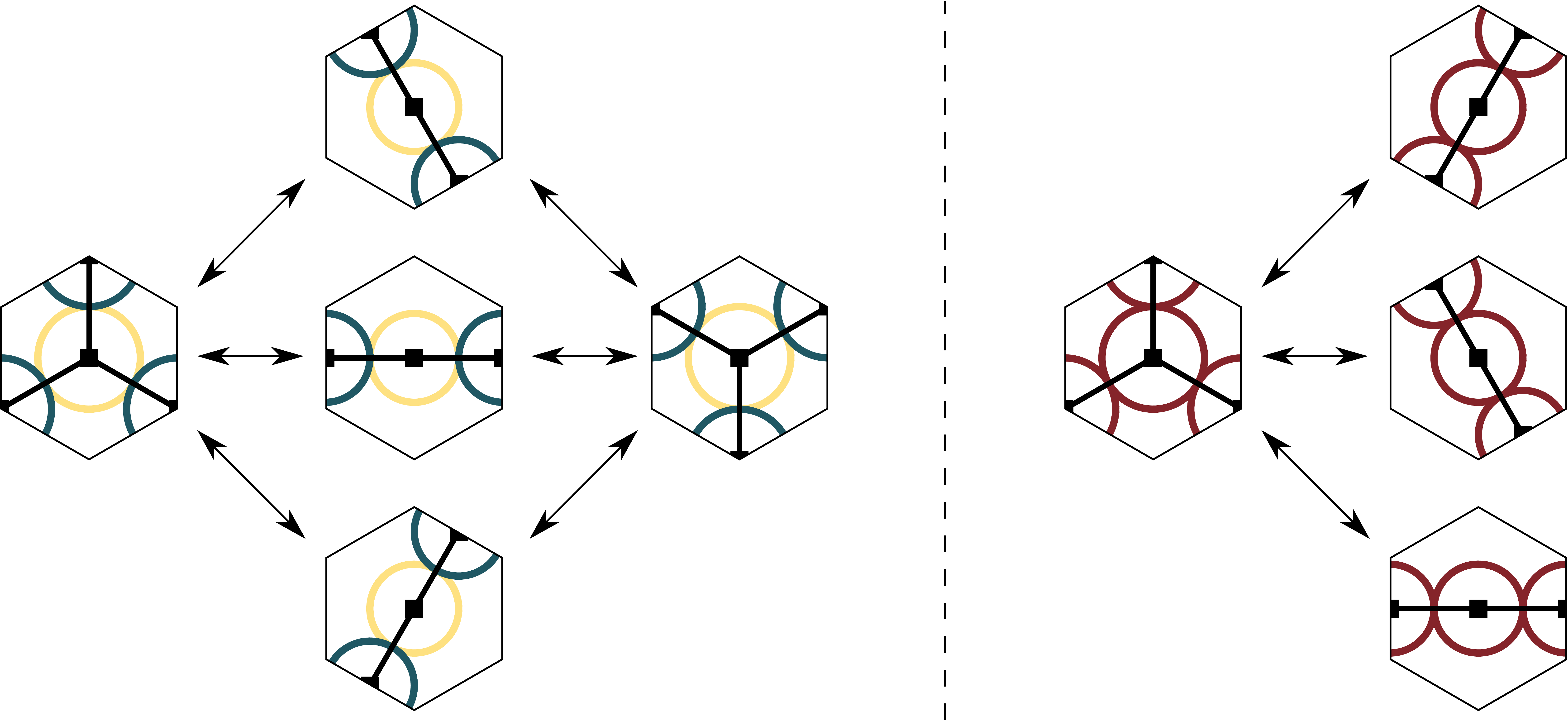}
		\caption{The critical points of the translation invariant configuration spaces of two hard disks on a hexagonal torus (labeled on the left, unlabeled on the right). The configurations where the disks have three connections are local minima, and those with two connections are saddle points.}
		\label{fig:figure1}
	\end{figure}

	The Topological Hypothesis effectively associates the topological changes indicated by critical points with geometric changes in the accessible region of the configuration space. The specific relationship of the topology to the geometry depends on how the configuration space is constructed though. Initially consider fixing a coordinate system to identify points in a spatial region $X$, assigning labels to each of $n$ particles, and representing every possible configuration of this system by a point in the product space $X^n$. This is not ideal from the standpoint of physically-distinguishable configurations though, e.g., a single configuration of identical particles with two distinct labelings corresponds to two distinct points in this configuration space. For specificity, consider the case of two hard disks in the hexagonal torus, with critical points as shown in Fig.\ \ref{fig:figure1}. For a disk radius $\rho$, the accessible region of configuration space is the connected component of configurations where every pair of disk centers is separated by at least $2 \rho$. The geometry of the accessible region can therefore be considered as a function of $\rho$, with the configuration space being empty in the large radius limit and the configuration space of points in the zero radius limit. As the disk radius decreases in the labeled configuration space on the left, two disconnected components initially appear. A system beginning in one of these sub-spaces cannot cross into the other unless the disk radius is further decreased, allowing the configuration to pass through one of the three saddle points. That is, the volume of the accessible region increases discontinuously at this disk radius. However, there is only ever one connected component in the unlabeled space on the right, making the volume of the accessible region a continuous function of disk radius. This has significant implications if the configurational entropy is defined as a function of the volume of the accessible region, since the configurational entropy would then be discontinuous on the left but not on the right.
	
	The canonical approach is to quotient the configuration space by all possible symmetries. For example, the homogeneity of space encourages the use of center of mass coordinates in classical mechanics \cite{denman1965invariance}. When constructing regression functions for the potential energy of local atomic environments, it has been reported \cite{behler2007,kocer2019} that using a configuration space that is invariant to translations, permutations and rotations decreases the number of training points required and increases the accuracy of the regression. However, the example in Fig.\ \ref{fig:figure1} suggests that quotienting by such symmetries could affect the geometry and topology of the configuration space in unexpected ways.

	The hard disk system is often considered as a prototype for simple fluids \cite{dyre2016simple}. It is governed by the hard disk potential, defined to be infinite if any pair of disk centers is separated by less than the sum of their radii and zero otherwise, and was first studied by Alder and Wainwright \cite{alder1962phase} almost sixty years ago. A number of studies suggest that the hard disk system undergoes at least one phase transition with varying packing fraction $\eta$ of the disks, with the solid and liquid phases perhaps separated by an intermediate hexatic phase. A solid characterized by long-range translational and orientational order is observed when $\eta > 0.72$, whereas a liquid characterized by the absence of any long-range order is observed when $\eta < 0.70$ \cite{mitus1997local,weber1995melting}. The behavior in the $0.70 < \eta < 0.72$ interval is a subject of ongoing controversy. This was initially believed to be a two-phase region exhibiting large fluctuations in density, generally considered as a sign of a first-order phase transition. Halperin, Nelson \cite{halperin1978theory} and Young \cite{young1979melting} instead suggested that the transition could be of Kosterlitz--Thouless type, implying the existence of a hexatic phase in this interval. Conflicting results continue to be reported in the literature about the order of the transition and the phases involved. Marx et al.\ \cite{mitus1997local,weber1995melting} reported a single step first order phase transition, whereas Bernard and Krauth \cite{bernard2011two} and Engel et al.\ \cite{engel2013hard} reported a two step phase transition with a first order liquid-hexatic transition and a second order solid-hexatic transition. Given this controversy, an approach that could identify the onset of a phase transition from more fundamental considerations than a discontinuous change in the value of a thermodynamic quantity could resolve the question of what happens in the $0.70 < \eta < 0.72$ interval, and would likely be useful in a broader thermodynamic context as well. While we do not claim to complete such an undertaking here, the necessary machinery is developed and a case study suggests that such an approach is in principle possible.
	
	Configuration spaces of hard disks have been studied previously \cite{carlsson2012harddisks,baryshnikov2014min}. Ritchey \cite{ritcheyphd} specifically studied the configuration spaces of hard disks on the hexagonal torus. They provided appropriate definitions of critical points and critical index, and examined the equivalence classes that critical points form under the action of translation, permutation and discrete lattice symmetries. A high density of critical points around the packing fraction of the solid-liquid transition indicated rapidly-changing configuration space topology there. This is suggestive of idea underlying the Topological Hypothesis, i.e., that a signature of two-dimensional hard disk melting should be visible in the distribution of critical points of the potential energy surface in the corresponding configuration space. One area not comprehensively addressed by this earlier work is the effect that quotienting out by the action of symmetry groups has on the number and distribution of the critical points.
	
	Roughly speaking, a quotient map sends a set of points in the base configuration space differing only by the action of a symmetry group to a single point in the quotient space. That is, the quotient map collects configurations with identical physical properties into equivalence classes, and the quotient space describes how the equivalence classes should be related to one another to preserve sensible notions of similarity. The equivalence classes studied by Ritchey \cite{ritcheyphd} effectively define a set of quotient maps and quotient spaces that are studied in more detail here.
	
	As far as the authors know, explicit triangulation of the configuration spaces of hard disks, quotiented by symmetry groups or otherwise, has never been done before. Our purpose here is to establish that this can be accomplished using topological data analysis techniques, and to show that the resulting triangulation allows study of the topological and geometric properties of the configuration spaces. The approach is demonstrated for the comparatively simple but nontrivial cases of two hard disks on the square and hexagonal toruses. While these should not be expected to resolve what happens in the $0.70 < \eta < 0.72$ interval of the hard disk system in the thermodynamic limit, the insights gained from these simpler systems are envisioned as part of a larger effort to develop a more precise formulation of the Topological Hypothesis, and eventually to evaluate whether such a hypothesis holds in practice.
	
	More specifically, this article is concerned with using explicit triangulations of the configuration space to study the action of quotient maps induced by symmetry groups on the number and distribution of critical points. Constructing explicit triangulations of the configuration space and the various quotient spaces is not trivial even for two disks, and is sufficient to demonstrate many of the same concerns that will likely arise for more complicated systems. Three quotient spaces of the base configuration space are considered. The first quotients out only the translational symmetry. The second adds the permutation symmetry of the disk labels and the inversion. The third adds the discrete symmetries of the lattice implied by the boundary conditions. Distance functions that respect the topology of the spaces and appropriately identify symmetry-related points are proposed, and are essential to the study of these spaces. Explicit triangulations are constructed using the $\alpha$-complex \cite{edelsbrunner1983shape}, and the isometric feature mapping (ISOMAP) algorithm \cite{Tenenbaum2319} is used for dimensionality reduction.
	
	Section \ref{sec:tautological_function} defines the configuration spaces of $n$ disks of radius $\rho$ using the tautological function. Section \ref{sec:morse_theory} briefly introduces concepts from classical Morse theory that are relevant to the discussion of critical points. Section \ref{sec:distance_config} provides unambiguous definition of the symmetry groups considered here, and proposes closely-related distance functions on the base configuration space and all of the quotient spaces. Section \ref{sec:descriptors} defines a procedure to map a hard disk configuration into a space with coordinates that are invariant to the desired symmetry groups. Finally, Sec.\ \ref{sec:configuration_spaces} presents and discusses the explicit triangulations of the quotient spaces as a function of disk radius.

	\section{Tautological function}
	\label{sec:tautological_function}

\begin{figure}
		\centering
		\includegraphics[width=0.85\columnwidth]{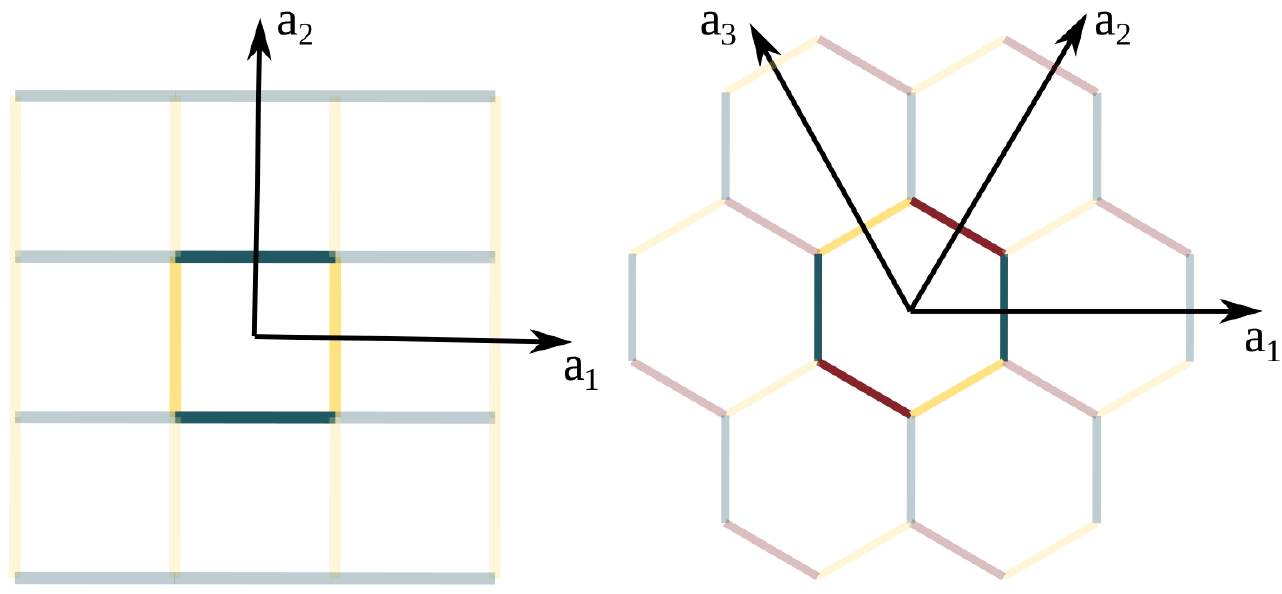}
		\caption{A torus is obtained by identifying the opposite edges of a square (left) or a hexagon (right). These can be lifted to tilings of the plane, with the fundamental cells containing the origin and the periodic images shown in faint outline. The center to center distance of neighboring cells is always one.}
		\label{fig:figure2}
	\end{figure}

	The configuration space of $n$ points on a torus $T^2$ is the product space of $n$ toruses, or 
	\begin{equation*}
	\Lambda(n) = \{ \config{x} = (\vec{x}_1, \dots ,\vec{x}_n) \;|\; \vec{x}_i \in T^2 \}.
	\label{eq:configuration_space1}
	\end{equation*}
	Figure \ref{fig:figure2} shows the square and hexagonal toruses used in this study; periodic boundary conditions are imposed by identifying opposite edges of both domains. Two domains are studied to help separate the specific and general phenomena that can occur when quotienting a configuration space by the action of a symmetry group. More generally, any numerical study of the Topological Hypothesis for the hard disk system will require a choice of domain, and it will be necessary to distinguish what are consequences of that choice and what are nascent features of the thermodynamic system.
	
	The tautological function $\tau: \Lambda(n) \rightarrow R$ is defined as
	\begin{equation*}
	\tau = \min\limits_{\substack{1 \leq i < j \leq n}} {r_{ij}}
	\label{eq:tautological_function}
	\end{equation*}
	where $r_{ij}$ is half the geodesic distance between the centers of disks $i$ and $j$. Intuitively, $\tau$ is the maximum radius that the disks could have without any pair overlapping given the positions of the disk centers. Observe that the configuration space
	\begin{equation}
	\Gamma(n,\rho) = \tau^{-1}[\rho,\infty)
	\label{eq:configuration_space2}
	\end{equation}
	of $n$ hard disks of radius $\rho$ is the superlevel set of $\tau$, or the set of all configurations that could accommodate disks of radius at least $\rho$.

	\section{Morse theory}
	\label{sec:morse_theory}
	
	Equation \ref{eq:configuration_space2} represents the configuration space of hard disks by means of the superlevel sets of $\tau$. This should allow a Morse-type theory to be used with the the critical points of $\tau$ to identify changes in the configuration space topology. Classical Morse theory \cite{morse1934,milnor2016morse} relates the topology of a manifold $M$ to the critical points of a generic smooth function $f$ defined on that manifold. A critical point is defined as a point where the gradient $\nabla f$ of the function vanishes, and has a critical index equal to the number of negative eigenvalues of the Hessian matrix there. Intuitively, the critical index is the number of independent ways that one could move to decrease the value of $f$ to second order.

	Let $M_a = \{x \in M \,|\, f(x) < a\}$ denote a sublevel set of $M$. The fundamental theorem of Morse theory states that the topology of $M_a$ and $M_b$ are the same if the interval $[a, b]$ doesn't contain a critical point. If it instead contains an index-$p$ critical point, then the topology of $M_a$ and $M_b$ differ in a way that is equivalent to attaching a $p$-handle to $M_a$; an $n$-dimensional $p$-handle is defined as a contractible smooth manifold $D^{p} \times D^{n-p}$ where $D^p$ is the $p$-dimensional disk. For example, a $0$-handle and a $2$-handle in two dimensions are both two dimensional disks $D^0 \times D^2$ and $D^2 \times D^0$ (though they are attached in different ways), whereas a $1$-handle is a rectangle $D^1 \times D^1$. The difficulty with this approach is that $\tau$ is not a smooth function, and in fact is not differentiable wherever the minimum disk separation is realized by more than one pair of disks. Our approach to handling this is explained elsewhere \cite{ritcheyphd}, but briefly, $\tau$ is replaced by a smooth function $E = \sum_{i < j} \exp[-w(r_{ij} - \rho)]$ that converges to the hard disk potential in the $w \rightarrow \infty$ limit. Moreover, there is a strictly monotone transformation of $E$ that converges to $\tau$ in the same limit, suggesting that the critical points of $\tau$ be identified with the limiting critical points of $E$.
	
\begin{figure}
		\centering
		\includegraphics[width=0.5\columnwidth]{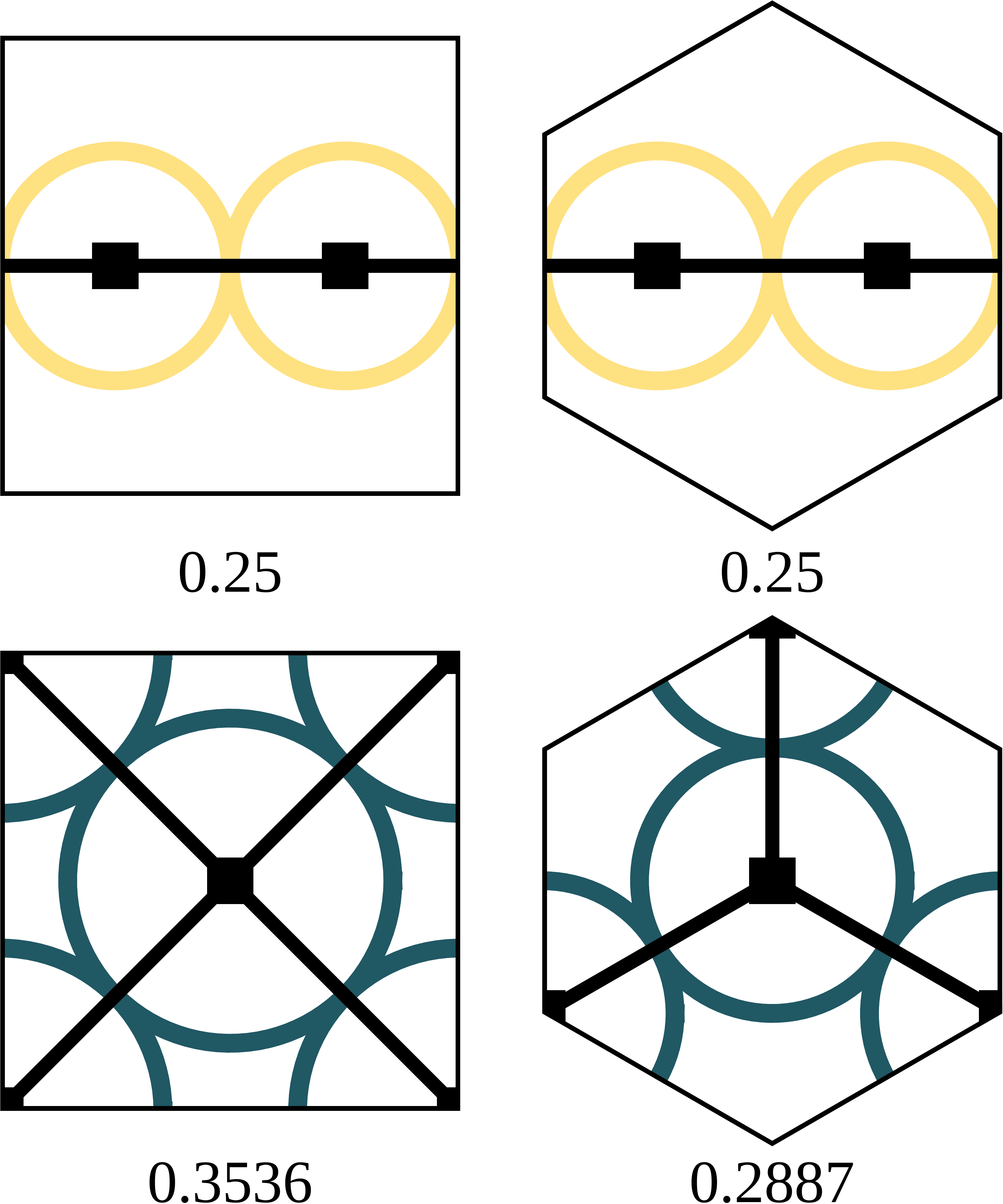}
		\caption{Representatives of the equivalence classes of critical points for two disks on the square and hexagonal toruses. The bottom (top) row corresponds to index-$0$ (index-$1$) critical points. The disk radius is reported below each configuration.}
		\label{fig:figure3}
	\end{figure}

	Practically, the critical points of $E$ for any finite $w$ can be found by searching for the minima of the scalar function $|\nabla E|^2$ using, e.g., the conjugate gradient algorithm. Initializing the algorithm with random configurations samples critical points with a weight that depends on the construction of $E$. The sampled critical points are grouped into equivalence classes containing configurations related by symmetry operations. Representatives of the equivalence classes found after millions of initializations for $n = 2$ are shown in Fig.\ \ref{fig:figure3}. Ritchey \cite{ritcheyphd} suggests that every critical configuration is reproduced infinitely many times by rigid translations (usually handled by fixing one of the disks at the origin), $n!$ times by permuting the disk labels, and some number of times related to the order of the plane tiling's symmetry group.
	
	\section{Distance}
	\label{sec:distance_config}
	
	The study of the configuration space geometry requires the definition of a suitable distance function. Depending on whether the space considered is the base configuration space or a quotient space, the distance could be defined between hard disk configurations or equivalence classes of configurations for given symmetry groups. For instance, the distance between two configurations that differ only by a translation should be nonzero in the base configuration space, but zero in the configuration space modulo translations where they belong to the same equivalence class.

\begin{figure}
		\centering
		\includegraphics[width=1.0\columnwidth]{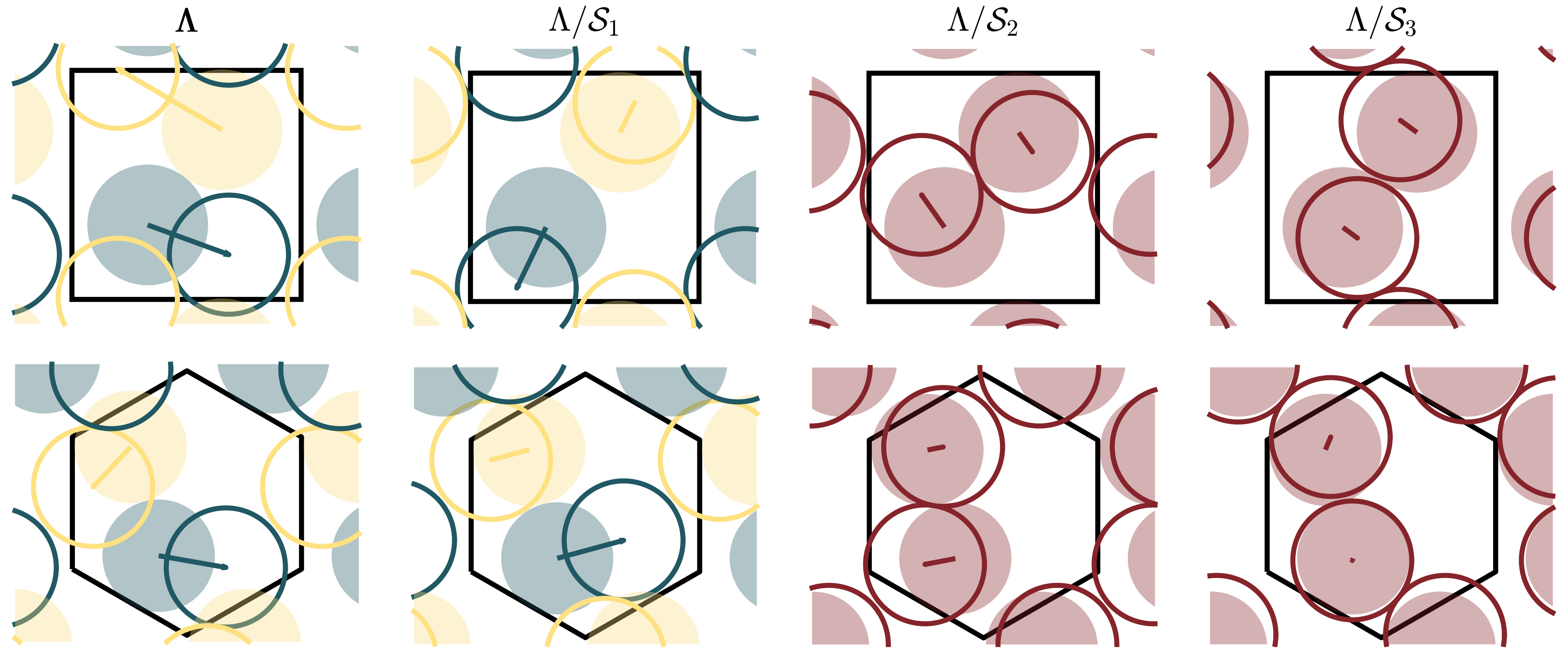}
		\caption{Distances between two configurations in the square and hexagonal toruses. The two configurations are indicated by filled and empty circles, and colors indicate the labelling of the disks. Table \ref{table:symmetries} lists the symmetry groups used to construct the quotient spaces.}
		\label{fig:figure4}
	\end{figure}

	One natural notion of distance assigns to two configurations $\config{p}, \config{q} \in \Lambda(n)$ a distance equal to the sum of the disk displacements required to transform one into the other, or
	\begin{equation}
	d_\Lambda(\config{p},\config{q}) = \sum_{i=1}^{n} {\Vert \vec{p}_i - \vec{q}_i \Vert}
	\label{eq:labeled_distance}
	\end{equation}
	where $\Vert \vec{p}_i - \vec{q}_i \Vert$ is the geodesic distance between the two positions of the $i$th disk. Figure \ref{fig:figure4} shows these displacements for two configurations sampled uniformly at random on the base configuration spaces for the square and hexagonal toruses. Here, $d_\Lambda$ is the sum of the lengths of the vectors pointing from one disk to the other. Observe that $d_\Lambda$ is sensitive to symmetry operations in the sense that applying translations, permutations or lattice symmetries to one of the configurations changes $d_\Lambda$. That said, $d_\Lambda$ satisfies the requirements of a metric on the base configuration space (identity of indiscernibles, symmetry, and the triangle inequality) with proofs provided in Appendix \ref{sec:metric_proofs}.
	
	The configuration space $\Lambda$ equipped with a metric $d_\Lambda$ constitutes a metric space $(\Lambda,d_\Lambda)$. Given a metric space and an equivalence relation $\sim$, there is a natural induced metric $d_{\Lambda / \sim}$ on the quotient space ${\Lambda/\!\!\sim}$ \cite{burago2001course}. When the equivalence relation additionally derives from a group of isometries $\mathcal{S}$, then the metric $d_{\Lambda / \mathcal{S}}$ on the quotient space $\Lambda / \mathcal{S}$ can be written as
	\begin{equation}
	d_{\Lambda / \mathcal{S}}(\config{p}, \config{q}) = \inf\limits_{\substack{S \in \mathcal{S}}} \{ d_\Lambda[\config{p}, S(\config{q})] \}.
	\label{eq:unlabeled_distance}
	\end{equation}
	Along with Eq.\ \ref{eq:labeled_distance}, this provides metrics on all the quotient spaces considered below.

	Let $\mathcal{T}$, $\mathcal{P}$, $\mathcal{I}$ and $\mathcal{L}$ respectively be the sets of rigid translations, permutations of the disk labels, inversion about the origin, and symmetries of the tiling of the plane. Formally, a configuration $\config{q}$ is a translation of $\config{p}$ by $\vec{t}$ if $\vec{q}_i = \vec{p}_i + \vec{t}$ for all disk indices $i$. Given a permutation $\pi \in \mathcal{P}$, $\config{q}$ is a permutation of $\config{p}$ if $\vec{q}_i = \pi(\vec{p}_i)$ for all $i$. A configuration $\config{q}$ is the inversion of $\config{p}$ if $\vec{q}_i = -\vec{p}_i$ for all $i$. Finally, for any symmetry element $L \in \mathcal{L}$ with representation $\mat{L}$, a configuration $\config{q}$ is a symmetric copy of $\config{p}$ if $\vec{q}_i = \mat{L} \vec{p}_i$ for all $i$. Observe that the operations belonging to all of these groups are isometric as required to use Eq.\ \ref{eq:unlabeled_distance}. Table \ref{table:symmetries} shows the different symmetry groups by which the configuration space is quotiented in this work, and the corresponding distances between the configurations in Fig.\ \ref{fig:figure4}. The $d_{\Lambda / S}$ are computed by fixing the first configuration and generating all copies of the second configuration that only differ by the action of $\mathcal{S / T}$, i.e., the discrete symmetry elements. Finding the rigid translation $T \in \mathcal{T}$ that minimizes $d_\Lambda\{\config{p}, T[S(\config{q})]\}$ for $S \in \mathcal{S / T}$ is a global optimization problem that is handled by the Tabu search algorithm \cite{chelouah2000tabu,glover1989tabu}. Finally, $d_{\Lambda / \mathcal{S}}$ is reported as the minimum of these distances for all $S \in \mathcal{S / T}$.
	
	\begin{table}
		\centering
		\begin{tabular}{||c c c c||} 
			\hline
			Space & Symmetries & $d_{\Lambda/\mathcal{S}}^{Square}$ & $d_{\Lambda/\mathcal{S}}^{Hexagon}$\\ [0.5ex] 
			\hline\hline
			$\Lambda$               & -             & 0.9048 & 0.5376 \\ 
			$\Lambda/\mathcal{S}_1$ & $\mathcal{T}$  & 0.4396 & 0.4739\\ 
			$\Lambda/\mathcal{S}_2$ & $\mathcal{T} \bigcup \mathcal{P} \bigcup \mathcal{I}$ & 0.2780 & 0.2048\\
			$\Lambda/\mathcal{S}_3$ & $\mathcal{T} \bigcup \mathcal{P} \bigcup \mathcal{I} \bigcup \mathcal{L}$ & 0.1687 & 0.0704\\[1ex]
			\hline
		\end{tabular}
		\caption{Isometric symmetry groups applied to the configuration space, and the corresponding distances between the configurations in Fig.\ \ref{fig:figure4}. $\mathcal{T}$, $\mathcal{P}$, $\mathcal{I}$ and $\mathcal{L}$ are the groups of translations, permutations, inversions, and symmetries of the tiling.}
		\label{table:symmetries}
	\end{table}
	
	The left column of Fig.\ \ref{fig:figure4} and the first row of Tab.\ \ref{table:symmetries} show the distance between configurations in the base configuration space $\Lambda$. The distance in $\Lambda/\mathcal{S}_1$ where $\mathcal{S}_1 = \mathcal{T}$ is the infimum of $d_\Lambda$ over all rigid translations of one configuration with respect to the other, including those that translate the disks across the edge of the fundamental cell. The distance in $\Lambda/\mathcal{S}_2$ where $\mathcal{S}_2 = \mathcal{T} \bigcup \mathcal{P} \bigcup \mathcal{I}$ is additionally minimized over permutations of the disk labels (indicated by the uniform disk color) and inversion about the origin. The distance in $\Lambda/\mathcal{S}_3$ where $\mathcal{S}_3 = \mathcal{T} \bigcup \mathcal{P} \bigcup \mathcal{I} \bigcup \mathcal{L}$ is additionally minimized over the symmetries of the tiling, i.e., the symmetries of the square and hexagon. Observe that the distance between two configuration cannot increase (and generally decreases) as more symmetries are included.

	\section{Descriptors}
	\label{sec:descriptors}
	
	As stated previously, the configuration space in Eq.\ \ref{eq:configuration_space2} contains redundant information. Specifically, every configuration is equivalent to multiple other configurations related by the symmetry operations discussed by Ritchey \cite{ritcheyphd}. As mentioned in the Introduction, quotienting the configuration space by the translation group is so widespread that this operation is often not explicitly mentioned. The motivation to do so is that the resulting configuration space is much smaller than the base configuration space. That said, the quotient maps are such that it is often not clear how to explicitly parameterize the quotient spaces, though this would certainly facilitate the construction of an explicit triangulation. This section describes our procedure to do so.

	Recall that the base configuration space for two disks is the product space $T^2 \times T^2$. Fixing the first disk at the origin effectively quotients the space by the translation group, making $\Lambda/\mathcal{S}_1$ equivalent to $T^2$. This is explicitly parameterized starting with a rectangular region with edge lengths $a$ and $b$ centered at the origin in the plane. The torus formed by identifying opposite edges of the rectangle has major radius $R = a / 2 \pi$ and minor radius $r = b / 2 \pi$. The coordinates of this torus in $R^3$ are given by
	\begin{align*}
	x' &= (R + r \cos\theta) \cos\phi \\
	y' &= (R + r \cos\theta) \sin\phi \\
	z' &= r \sin\theta
	\end{align*}
	where $\phi = (a / 2 - x) / R$ and $\theta = (b / 2 - y) / r$. This is used for the visualizations of $\Lambda/\mathcal{S}_1$ below.
	
	All other quotient spaces are initially embedded in an infinite-dimensional descriptor space, and a numerical approach is used to estimate the minimum number of descriptors necessary to maintain the embedding. Given a configuration space of $n$ disks, the distribution $f$ is defined as a sum of Dirac-delta distributions $\delta(\vec{a}_j)$ located at the disk centers $\vec{a}_j$ in the $a_1a_2$-coordinate system in Fig.\ \ref{fig:figure2}, or
	\begin{equation}
	f(\vec{a}) = \sum_{j = 1}^{n} \delta(\vec{a}_j) = \sum_{\vec{k}} {c_{\vec{k}} e^{2\pi i \vec{k} \cdot \vec{a}}}
	\label{eq:f}
	\end{equation}
	where $c_{\vec{k}}$ are the complex coefficients of the reciprocal lattice expansion and $\vec{k} = [p, q]$ for integers $p$ and $q$. The infinite set of $c_{\vec{k}}$ can be calculated using the orthogonality of the complex exponentials as
	\begin{equation}
	c_{\vec{k}} = \sum_{j=1}^{n} e^{-2\pi i \vec{k} \cdot \vec{a}_j} .
	\label{eq:c_k}
	\end{equation}
	$c_{\vec{k}}$ respects the periodicity of the lattice and is invariant to permutations of the disk labels due to the commutative property of the summation in Eq.\ \ref{eq:f}. It can be shown that translating a configuration (by adding an offset to the $\vec{a}_j$) only changes the phase of the coefficients. This means that the moduli of the coefficients, or
	\begin{equation}
	z_{\vec{k}} = \sqrt{c^\ast_{\vec{k}} c_{\vec{k}}}
	\label{eq:z_k}
	\end{equation}
	where $^\ast$ denotes the complex conjugate, are a set of real-valued descriptors that are invariant to disk label permutations and rigid translations. Observe that the descriptors $z_{\vec{k}}$ also respect inversion symmetry. Numerical experiments indicate that the rank of the Jacobian of the map from the $\vec{a}_j$ to the $z_{\vec{k}}$ is generically $2(n - 1)$, suggesting that some number of these descriptors could be sufficient to construct an embedding of $\Lambda / \mathcal{S}_2$.
	
	Constructing an embedding of $\Lambda / \mathcal{S}_3$ further requires the descriptors to be invariant to the symmetries of the plane tiling. This is done explicitly as
	\begin{equation}
	\hat{z}_{\vec{k}} = \frac{1}{O(\mathcal{L})}\sum_{L \in \mathcal{L}} z_{\vec{k}}^{L}
	\label{eq:z_hat_k}
	\end{equation}
	where $z_{\vec{k}}^{L}$ are the descriptors $z_{\vec{k}}$ of the configuration $L \config{x}$, i.e., a copy of $\config{x}$ acted upon by the symmetry operation $L \in \mathcal{L}$, and $O(\cdot)$ is the order of a group.
	
	Appendix \ref{sec:descriptor_proofs} provides a proof that not all of these descriptors are independent. The invariance of the descriptors $z_{\vec{k}}$ to the inversion implies that the descriptors for indices $\vec{k}$ and $-\vec{k}$ of a given configuration are the same for both the square and the hexagonal domains. The invariance of the descriptors $\hat{z}_{\vec{k}}$ to the symmetries of the plane tiling results in more complicated relationships that are fully described in Appendix \ref{sec:descriptor_proofs}. The set of independent descriptors closest to the origin in reciprocal space is always used in the analysis below.
	
	The maps into the infinite-dimensional spaces of descriptors are conjectured to be injective, i.e., to contain all information about the original configuration up to the desired symmetries. Since the number of disks is finite, it is likely that a finite number of dimensions (descriptors) is sufficient for this purpose though. The challenge then is to find the minimum number of descriptors necessary to maintain a proper embedding. The strategy proposed here is to order the descriptors by distance from the origin in reciprocal space, sequentially remove any dependent descriptors, and numerically search for self-intersections of the image space as a function of the number of descriptors retained after truncation.
	
\begin{figure}
		\centering
		\includegraphics[width=0.8\columnwidth]{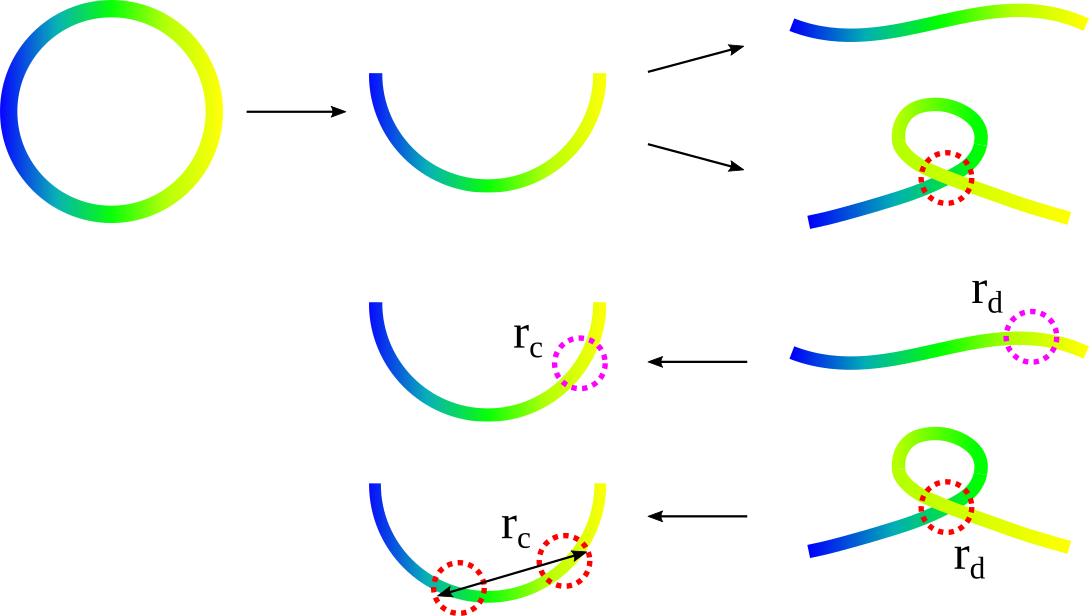}
		\caption{An illustration of the self-intersection search. The full circle on the left represents the base configuration space, with points related by a symmetry operation in the same color. The middle half-circle represents the space quotiented by the symmetry group, and on the right are possible images of the map into a truncated descriptor space. One of these preserves the embedding, but the one that self-intersects (indicated by the red dotted circle) does not. The self-intersection is identified by considering the diameter of the preimage of a neighborhood around the intersection.}
		\label{fig:figure5}
	\end{figure}

	Figure \ref{fig:figure5} illustrates the idea underlying the search for self-intersections. The full circle on the left represents the base configuration space, with points related by a symmetry operation in the same color. Quotienting by the symmetry group (folding the top half of the circle onto the bottom half) gives the quotient space represented by the half circle in the middle. On the right are possible images of the map of the quotient space into the truncated descriptor space. The number of descriptors could be sufficient for the image to be an embedding, as represented on the top right. The image could be self-intersecting if the number of descriptors is not sufficient though, as indicated by the region in the red dashed circle. The search for self intersections therefore involves sampling neighborhoods of radius $r_d$ in the descriptor space and examining the preimages of these neighborhoods. If the radius $r_c$ of the preimage scales with $r_d$ for all such neighborhoods, then the map into the descriptor space is likely an embedding. If $r_c$ appears to be independent of $r_d$ for any neighborhood, then this is likely due to $r_c$ measuring the distance between distinct neighborhoods in the preimage.
	
\begin{figure}
		\centering
		\includegraphics[width=1.0\columnwidth]{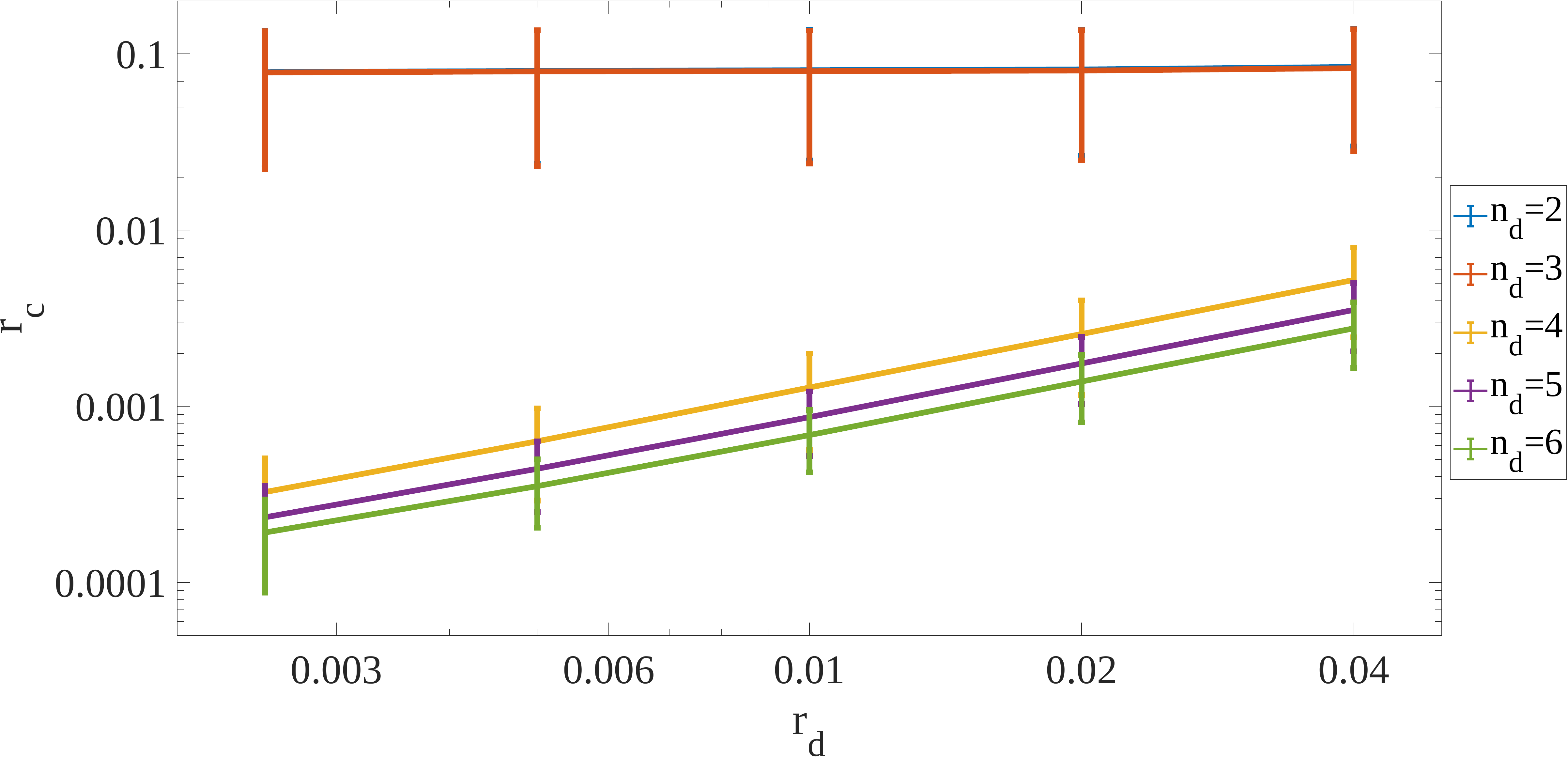}
		\caption{The inverse analysis for $n_d = 2 \dots 6$ with different $r_d$ values. The mean and standard deviation of $r_c$ approach zero as $r_d$ decreases for $n_d \ge 4$, suggesting that $n_d = 4$ is sufficient to embed the quotient space $\Lambda/\mathcal{S}_2$.}
		\label{fig:figure6}
	\end{figure}

	Practically, the procedure begins by sampling $N$ configurations uniformly at random in the base configuration space. For each of these configurations, the first $n_d$ descriptors that are invariant to the desired symmetries are computed. Small neighborhoods of radius $r_d$ are then defined about the images of each configuration in the descriptor space; suppose that $N_n$ images of other configurations lie within a particular neighborhood. The distances as defined in Sec.\ \ref{sec:distance_config} are computed between these $N_n$ configurations and the central configuration, and are used to estimate the radius $r_c$ of the preimage in the quotient space. If $r_c$ goes to zero as $r_d$ goes to zero for every neighborhood in the image, then the quotient space is likely embedded in the descriptor space. If not, then the image of the quotient space is likely self-intersecting as shown in Fig. \ref{fig:figure5}, $n_d$ is increased by one, and the process is repeated. Figure \ref{fig:figure6} shows the results of this analysis for the quotient space $\Lambda/\mathcal{S}_2$ and $n_d = 2 \dots 6$. It clearly shows that the mean and standard deviations of $r_c$ go to zero as $r_d$ goes to zero for $n_d \geq 4$, but not for $n_d \leq 3$. We conclude that four descriptors are sufficient to embed the quotient space $\Lambda/\mathcal{S}_2$.
	
	\section{Configuration spaces}
	\label{sec:configuration_spaces}
	
	The map of the quotient space into the descriptor space can be viewed as a coordinate transformation, and the Jacobian matrix of the transformation can be found. The rank of this matrix gives the dimension of the resultant manifold at the point of evaluation \cite{krantz2012implicit}. Repeated sampling of the Jacobian matrix for the quotient space $\Lambda/\mathcal{S}_2$ and $n = 2$ disks suggests that the rank is generically two and that the image in the descriptor space is locally a $2$-manifold. However, Fig.\ \ref{fig:figure6} suggests that at least four descriptors are required for the image in the descriptor space to be an embedding. Various dimensionality-reduction techniques can be used to try to reduce this further, enough to be able to visualize the space; the ISOMAP algorithm \cite{Tenenbaum2319} is used here. Intuitively, this algorithm attempts to find a lower-dimensional embedding that preserves the geodesic distances of the points in $k$-nearest neighbor graphs.

	Sampling hard disk configurations uniformly at random in the base configuration space and then computing the appropriate descriptors gives a point cloud embedded in the truncated descriptor space. The study of the topological and geometric properties of the quotient space would be significantly simpler with a simplicial complex instead of a point cloud though. While there are a variety of simplicial complexes used in the literature on statistical topology (e.g., the Vietoris--Rips \cite{vietoris1927hoheren} and Cech \cite{hatcher2002algebraic} complexes), this work uses the $\alpha$-complex \cite{edelsbrunner1983shape} which is a subcomplex of the Delaunay triangulation \cite{delaunay1934sphere}. Formally, let $P$ be a set of points in $R^d$ and $\Delta_k$ be a $k$-simplex where $0 \leq k \leq d$. Let $r$ and $c$ be the radius and the center of the circumsphere of $\Delta_k$, respectively. Given the Delaunay triangulation $DT(P)$ of $P \subset R^d$, the $\alpha$-complex $C_\alpha(P)$ of $P$ is a simplicial subcomplex of $DT(P)$ such that a simplex $\Delta_k \in DT(P)$ is in $C_\alpha(P)$ if (i) $r<\alpha$ and the $r$-ball located at $c$ is empty, or (ii) $\Delta_k$ is a face of another simplex in $C_\alpha(P)$.
	
\begin{figure}
		\centering
		\includegraphics[width=1.0\columnwidth]{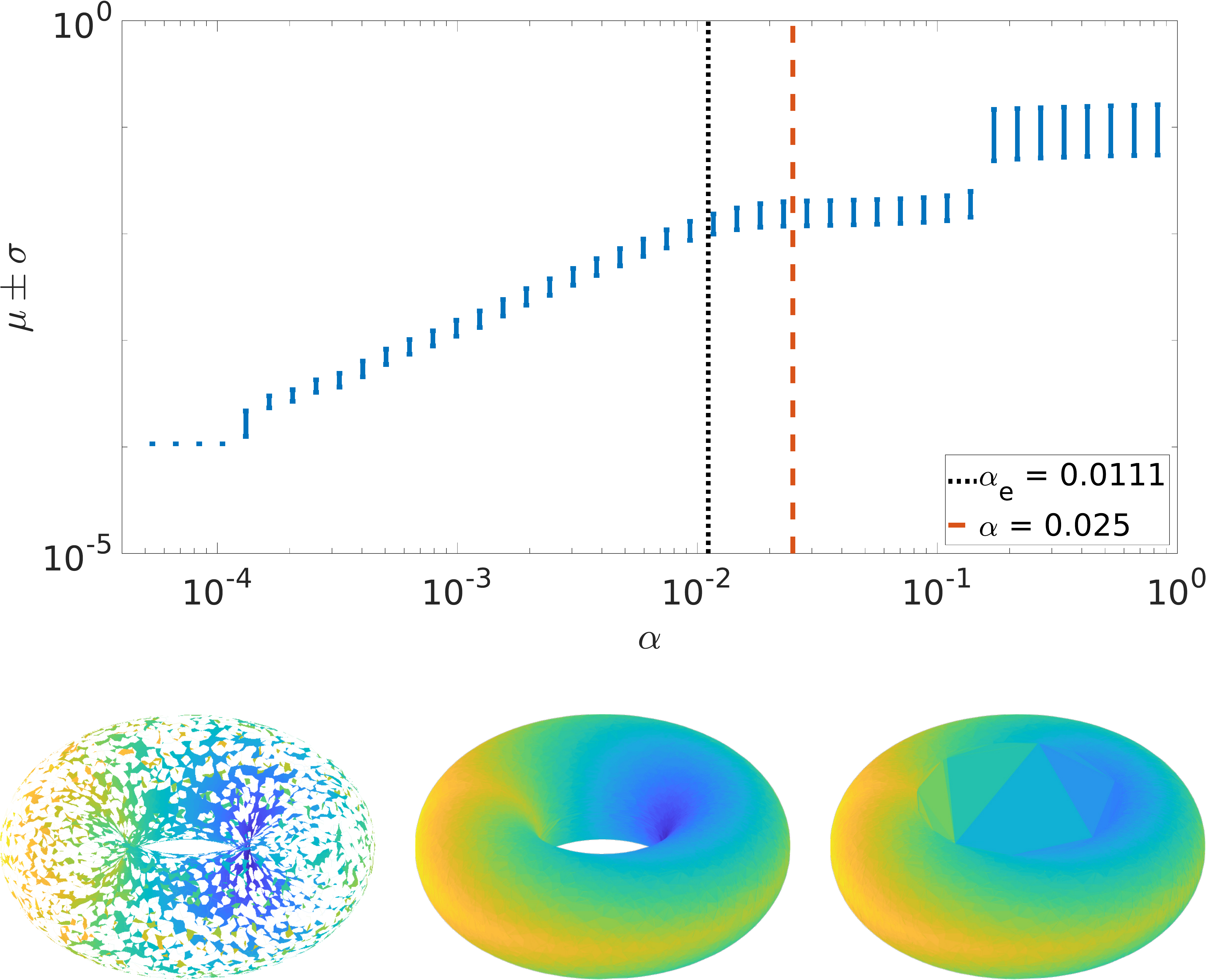}
		\caption{The length scale analysis for $\Lambda/\mathcal{S}_1$ for the square torus. $\mu$ and $\sigma$ denote the mean and standard deviation of the edge lengths of the $\alpha$-complex. The black-dotted line shows the lower bound estimate $\alpha_e$, whereas the red-dashed line shows the $\alpha$ value actually used to construct the $\alpha$-complex. $\alpha$-complexes for increasing $\alpha$ values $\{0.0001, 0.025, 0.5\}$ are shown at the bottom.}
		\label{fig:figure7}
	\end{figure}

	A persistent question with $\alpha$-complexes is the appropriate value of $\alpha$. Our intention is to find a value such that the $\alpha$-complex in the truncated descriptor space is a reasonable approximation of the quotient space. The heuristic used here involves a length scale analysis of the edges in the complex as a function of $\alpha$. Let $\mu$ and $\sigma$ respectively be the mean and standard deviation of the edge lengths. For very small $\alpha$ values, the $\alpha$-complex contains only $0$-simplices and a few $1$-simplices and $\mu$ and $\sigma$ are very small. For large $\alpha$ values, the $\alpha$-complex approaches the full Delaunay triangulation, simplices that connect distant points are included, and $\mu$ and $\sigma$ are large. For intermediate $\alpha$ values, there is presumably a plateau with intermediate values of $\mu$ and $\sigma$ where the geometry of the complex is relatively stable (though this depends on the density of the sampled points). Any $\alpha$ within this plateau should be a reasonable value. An alternative would be to calculate the persistent homology as a function of $\alpha$ \cite{edelsbrunner2008persistent}, but this would probably not provide significantly different values from the simpler length scale analysis used here. Figure \ref{fig:figure7} shows the result of this length scale analysis for the quotient space $\Lambda/\mathcal{S}_1$, and suggests that $\alpha = 0.025$ is a reasonable value.
	
	A lower bound on $\alpha$ is estimated as follows. Given $n_p$ points in $d$ dimensions, the Delaunay triangulation contains $O(n_p^{d/2})$ simplices \cite{seidel1995upper}. This study always samples $n_p = 10^4$ points, giving $n_t \approx 10^6$ tetrahedra in the full Delaunay triangulation of a $2$-manifold embedded in $R^3$. Assuming that the volume of the convex hull of $\Lambda/\mathcal{S}_1$ for two disks is covered by uniform equilateral tetrahedra would give $\alpha_e = 2^{1/6} \times (6V/n_t)^{1/3}$ for the tetrahedral edge length where $V$ is the manifold's volume. Since the space for $\Lambda/\mathcal{S}_1$ is constructed using the rectangle $[0, 1] \times [0, 2]$, the lower bound is $\alpha_e = 0.0111$. As seen in Fig.\ \ref{fig:figure7}, this estimate is conservative.
	
	\subsection{Adding translation invariance}
	\label{subsec:space_translation}
	
\begin{figure}
		\centering
		\includegraphics[width=0.8\columnwidth]{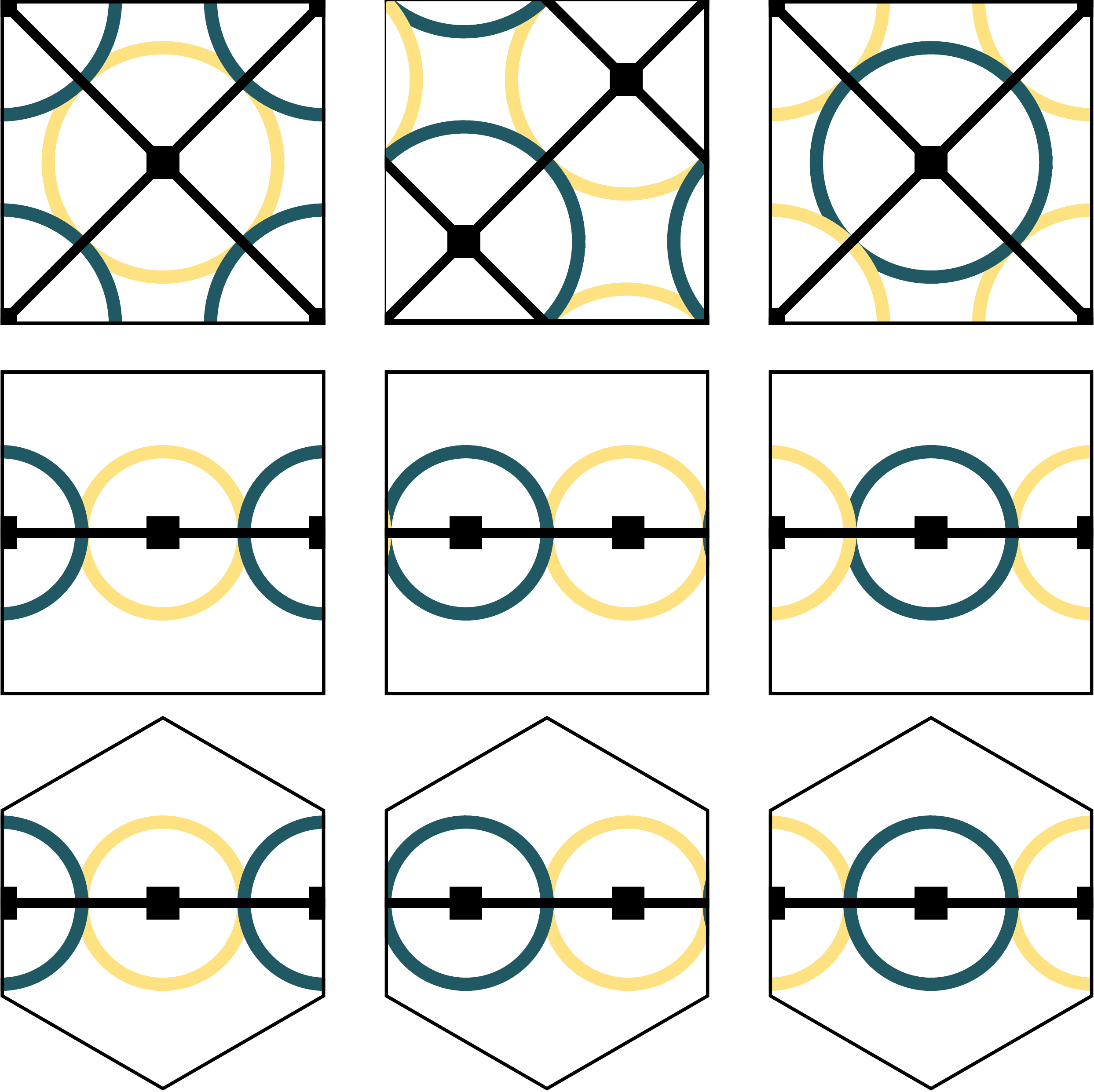}
		\caption{Critical points can be related by both translational and permutation symmetries. This happens for both critical points of the square, but only for the index-$1$ critical point of the hexagon.}
		\label{fig:figure8}
	\end{figure}

	The base configuration space $\Lambda$ with the function $\tau$ is not amenable to Morse theory since the critical points of $\tau$ are not isolated; in fact, every critical point is related by a rigid translation to an entire critical submanifold. Partly for this reason the usual practice is to quotient out the rigid translations by, e.g., fixing the position of the first disk. This apparently innocuous operation can have the unexpected effect of identifying points related by a permutation of the disk labels though. For example, consider the index-$0$ critical point in the top row of Fig.\ \ref{fig:figure8}. Translating the disks diagonally by the translation vector $\vec{t} = [0.5, 0.5]$ is equivalent to exchanging the disk labels, but is identified with the critical point on the left in the quotient space $\Lambda/\mathcal{S}_1$. Likewise, translating the index-$1$ critical point in the middle row to the right by $\vec{t} = [0.5, 0]$ is equivalent to exchanging the disk labels. That is, the submanifold that is identified when quotienting out by rigid translations can contain multiple points related by permutation symmetries. This implies that not all the equivalence classes of points related by permutation symmetries in $\Lambda/\mathcal{S}_1$ contain $n!$ elements, despite this being widely assumed (perhaps because each of these equivalence classes does contain $n!$ elements in $\Lambda$). Moreover, changing the domain of an integral from $\Lambda/\mathcal{S}_1$ to $\Lambda/\mathcal{S}_2$ is not generally as simple as dividing by a factor of $2 n!$ (the factor of $2$ for the inversion and $n!$ for the permutation group), despite this being standard practice in statistical mechanics \cite{gould2010statistical,peliti2011statistical}.
	
	\begin{figure*}
		\centering
		\includegraphics[width=1.0\textwidth]{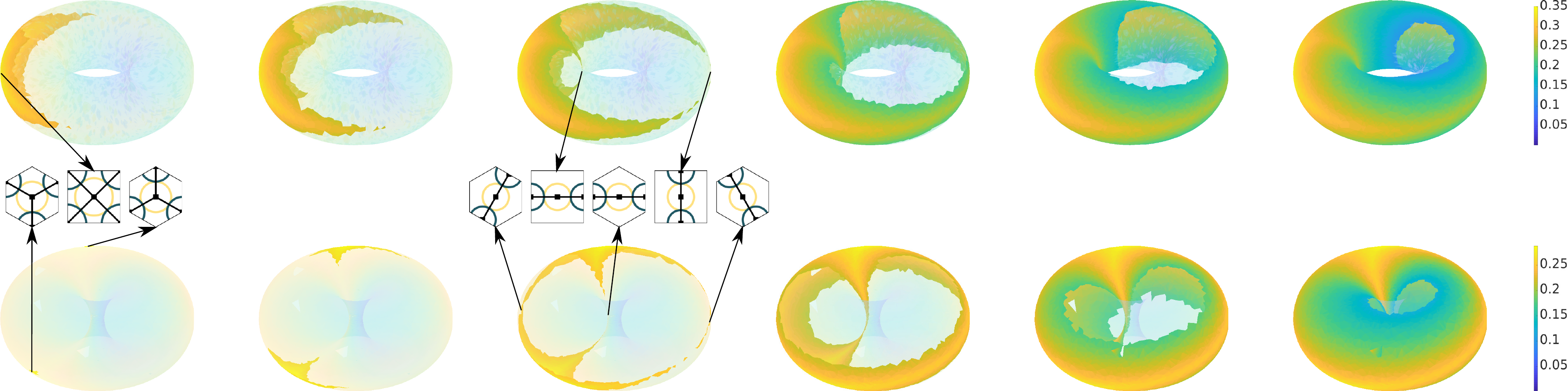}
		\caption{The evolution of the translation invariant configuration space $\Gamma(2, \rho) / \mathcal{S}_1$ for the square torus (top) and for the hexagonal torus (bottom) with $\rho=\{0.28, 0.26, 0.25, 0.21, 0.17, 0.12\}$. The locations of the critical points in this space are indicated by arrows.}
		\label{fig:figure9}
	\end{figure*}
	
	Figure \ref{fig:figure9} shows the translation-invariant configuration space $\Gamma(2, \rho) / \mathcal{S}_1$ of two disks as a function of $\rho$ for the square torus (top) and hexagonal torus (bottom) as obtained from the $\alpha$-complex of 10 000 points. Note that the square torus is constructed by extending the square to a rectangle and identifying opposite edges, but this does not affect the topological properties of the space. When $\rho > 0.25$, the space $\Gamma(2, \rho) / \mathcal{S}_1$ is comprised of a single $0$-handle whereas that of the hexagonal torus is comprised of two $0$-handles. This difference should be expected on the basis of Fig.\ \ref{fig:figure8} since the two index-$0$ critical points of the hexagonal torus are not related by a rigid translation. When $\rho = 0.25$, two and three $1$-handles are connected for the square and the hexagonal toruses, respectively. Observe that the $1$-handles provide connections between previously distant regions of the space. For $\rho < 0.25$, the space continues to grow and eventually closes in the $\rho \rightarrow 0$ limit. That is, the configuration with $\rho = 0$ acts like an index-$2$ critical point, even though it is not strictly within the space.

	Figure \ref{fig:figure9} further confirms that some critical points are related by both translation and permutation symmetries, since the numbers of index-$0$ and index-$1$ critical points are, e.g., $1$ and $2$ instead of the $2$ and $4$ expected for the square torus on the basis of the symmetry group orders. Finally, the topology of $\Lambda / \mathcal{S}_1$ is that of a torus for both the square and the hexagon, as expected.

	\subsection{Adding permutation and inversion invariance}
	\label{subsec:space_translation_permutation_inversion}
	
\begin{figure*}
		\centering
		\includegraphics[width=1.0\textwidth]{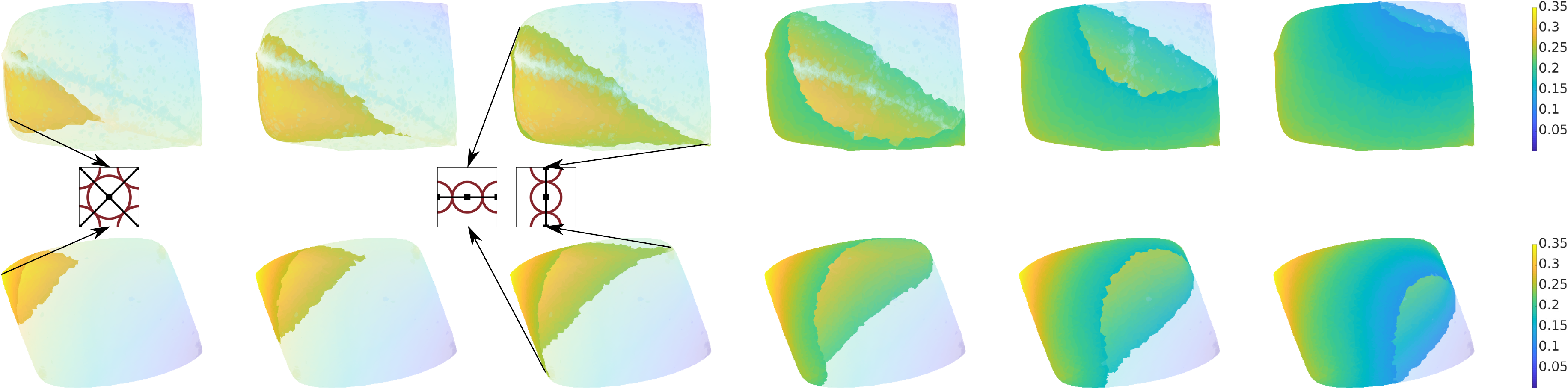}
		\caption{The evolution of the translation, permutation and inversion invariant configuration space $\Gamma(2, \rho) / \mathcal{S}_2$ for the square torus constructed with the standard Euclidean distance (top) and the distance in Eq.\ \ref{eq:unlabeled_distance} (bottom) with $\rho = \{0.28, 0.26, 0.25, 0.21, 0.17, 0.12\}$. The locations of the critical points are indicated by arrows.}
		\label{fig:figure10}
	\end{figure*}

\begin{figure*}
		\centering
		\includegraphics[width=1.0\textwidth]{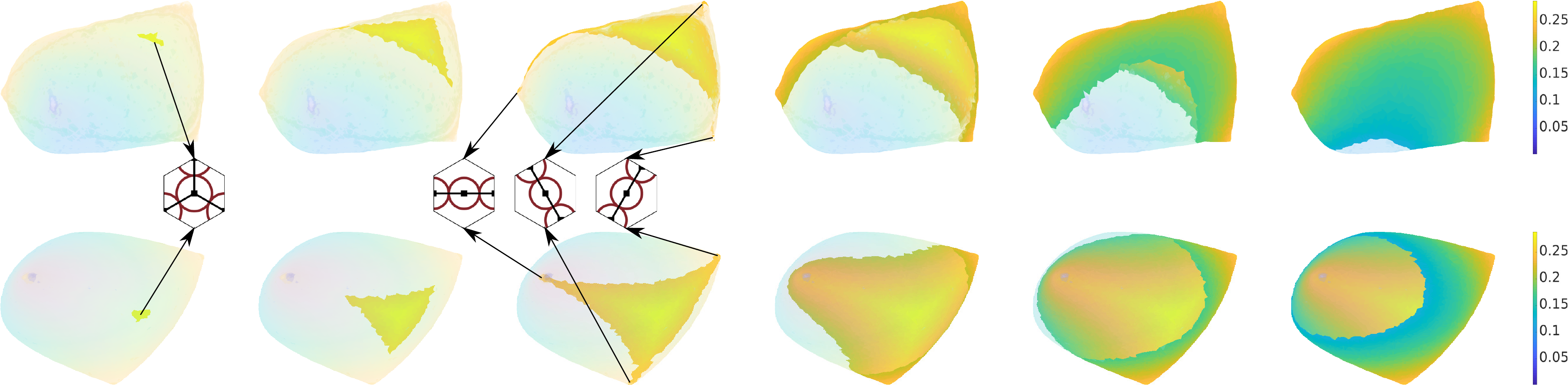}
		\caption{The evolution of the translation, permutation and inversion invariant configuration space $\Gamma(2, \rho) / \mathcal{S}_2$ for the hexagonal torus constructed with the standard Euclidean distance (top) and the distance in Eq.\ \ref{eq:unlabeled_distance} (bottom) with $\rho = \{0.28, 0.26, 0.25, 0.21, 0.17, 0.12\}$. The locations of the critical points are indicated by arrows.}
		\label{fig:figure11}
	\end{figure*}

	The descriptors $z_{\vec{k}}$ defined in Sec.\ \ref{sec:descriptors} are by construction invariant to rigid translations, inversions about the origin, and permutations of disk labels. One way to construct the quotient space $\Lambda / \mathcal{S}_2$ is then to use the $z_{\vec{k}}$ as coordinates for the descriptor space. Figure \ref{fig:figure6} suggests that four of these are sufficient for a proper embedding of $\Lambda / \mathcal{S}_2$. The ISOMAP algorithm is used to reduce the dimension further by one, allowing visualization of the quotient space, but requires a distance function to do so. The top rows of Fig.\ \ref{fig:figure10} and Fig.\ \ref{fig:figure11} use the Euclidean distance in the descriptor space, whereas the bottom rows use the distance defined in Eq.\ \ref{eq:unlabeled_distance}. This allows two versions of $\Gamma(2, \rho) / \mathcal{S}_2$ to be constructed for both the square and hexagonal toruses; it is significant that the two versions are topologically identical, though the one using Eq.\ \ref{eq:unlabeled_distance} better preserves the expected quotient space symmetries; analogous to the truncation of a Fourier series, the use of a distance based on a finite number of descriptors likely introduces distortions. Regardless, $\Gamma(2, \rho) / \mathcal{S}_2$ starts with the index-$0$ critical points and grows without topological change until $\rho = 0.25$ when the index-$1$ critical points appear. Unlike for $\Gamma(2, \rho) / \mathcal{S}_1$, these critical points don't appear as handles, but as singular points. 

	That critical points of the base configuration space do not behave in the same way in the quotient spaces should be emphasized; the index-$1$ critical points in Fig.\ \ref{fig:figure3} do appear in the $\Gamma(2, \rho) / \mathcal{S}_2$, but without any change in the topology. Instead, the critical points correspond to the appearance of sharp corners such that $\Gamma(2, 0.25) / \mathcal{S}_2$ cannot be described as a smooth manifold with boundary, but rather is a Whitney stratified space. Finally, that the critical points do not connect distant regions of the space significantly affects certain geometric properties, e.g., the diameter of the space as measured by the diffusion distance \cite{coifman2006diffusion}. As $\rho$ is further decreased, the spaces continue to grow and eventually close up, indicating that the topology of the quotient space $\Lambda / \mathcal{S}_2$ is that of a sphere rather than a torus. That all of these changes occurred when merely quotienting out by permutations of the disk labels suggests that the ideas motivating the Topological Hypothesis need to be explored with great care.
	
	\subsection{Adding lattice invariance}
	\label{subsec:space_translation_permutation_inversion_rotation}
	
\begin{figure*}
		\centering
		\includegraphics[width=1.0\textwidth]{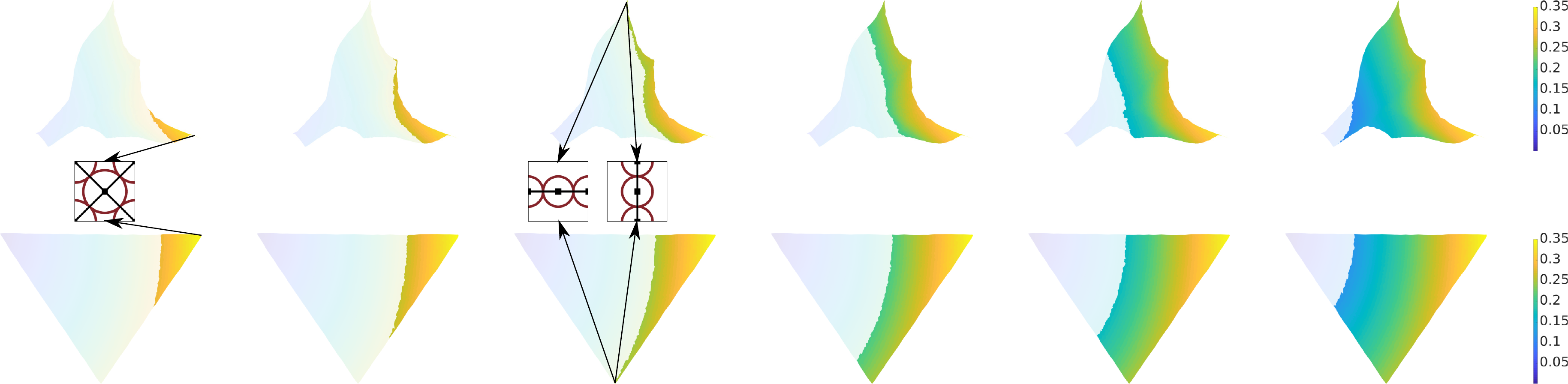}
		\caption{The evolution of the translation, permutation, inversion and lattice symmetry invariant configuration space $\Gamma(2, \rho) / \mathcal{S}_3$ for the square torus constructed with the standard Euclidean distance (top) and with the distance in Eq.\ \ref{eq:unlabeled_distance} (bottom) with $\rho=\{0.28, 0.26, 0.25, 0.21, 0.17, 0.12\}$. The locations of the critical points are indicated by arrows.}
		\label{fig:figure12}
	\end{figure*}

\begin{figure*}
		\centering
		\includegraphics[width=1.0\textwidth]{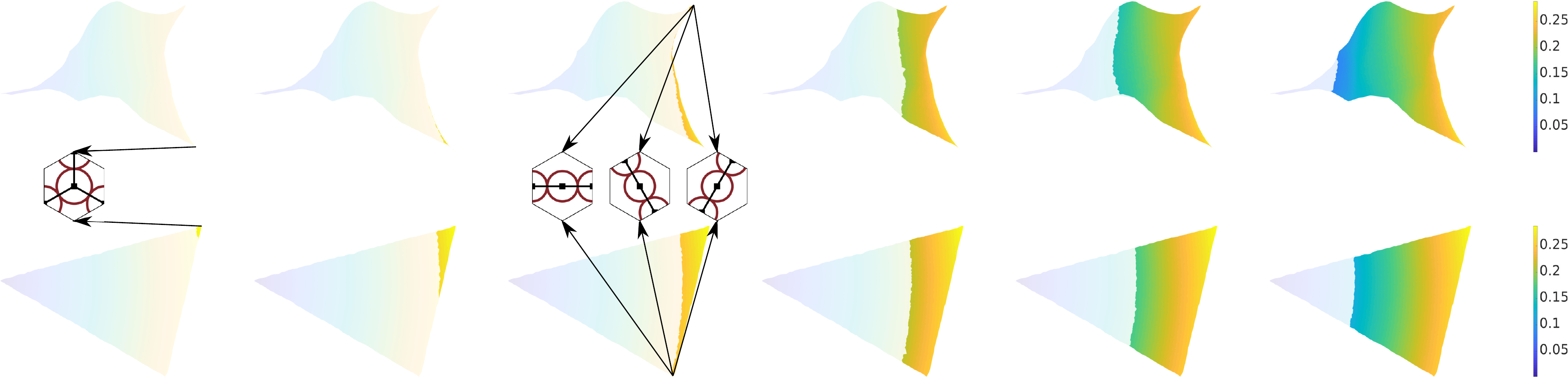}
		\caption{The evolution of the translation, permutation, inversion and lattice symmetry invariant configuration space $\Gamma(2, \rho) / \mathcal{S}_3$ for the hexagonal torus constructed with the standard Euclidean distance (top) and with the distance in Eq.\ \ref{eq:unlabeled_distance} (bottom) with $\rho=\{0.28, 0.26, 0.25, 0.21, 0.17, 0.12\}$. The locations of the critical points are indicated by arrows.}
		\label{fig:figure13}
	\end{figure*}

	The descriptors $\hat{z}_{\vec{k}}$ defined in Sec.\ \ref{sec:descriptors} are additionally invariant to the symmetries of the plane tiling, and are used as coordinates for the embedding of the quotient space $\Lambda / \mathcal{S}_3$. As before, dimensionality reduction is performed with the ISOMAP algorithm. The top rows of Fig.\ \ref{fig:figure12} and Fig.\ \ref{fig:figure13} use the Euclidean distance in the descriptor space, whereas the bottom rows use the distance defined in Eq.\ \ref{eq:unlabeled_distance}. The two versions of $\Gamma(2, \rho) / \mathcal{S}_3$ are topologically identical as before. That said, the one using Eq.\ \ref{eq:unlabeled_distance} better preserves the expected quotient space symmetries, with the geometric distortions introduced by using the Euclidean distance in the descriptor space much more pronounced than those in Fig.\ \ref{fig:figure10} and Fig.\ \ref{fig:figure11}. Specifically, the version of $\Gamma(2, \rho) / \mathcal{S}_3$ constructed with the Euclidean distance incorrectly collapses the region for small $\rho$ to a $1$-manifold. Further examination suggests that the quotient spaces constructed with Eq.\ \ref{eq:unlabeled_distance} are the smallest symmetric regions of their corresponding domains; the bottom row of Fig.\ \ref{fig:figure12} is $1 / 8$ of the square torus, whereas that of Fig.\ \ref{fig:figure13} is $1 / 12$ of the hexagonal torus. The corresponding fundamental cells can be obtained by reflecting the quotient spaces along an edge passing through the $\rho = 0$ point and applying the appropriate rotations.

	Observe that the topology of the quotient space is completely changed by quotienting out the symmetries of the plane tiling. The index-$0$ critical point doesn't correspond to a $0$-handle anymore, but to a single point, and the index-$1$ critical points are all identified by the symmetry operations. The $\rho = 0$ point appears as a single point as well, rather than as a $2$-handle as in the other quotient spaces considered here. Finally, $\Lambda / \mathcal{S}_3$ has a boundary and is topologically equivalent to a disk, in contrast to $\Lambda / \mathcal{S}_2$ having the topology of a sphere and $\Lambda / \mathcal{S}_1$ that of a torus.
	
	\section{Conclusion}
	\label{sec:conclusion}	
	
	The configuration space is essential to the statistical mechanics of glass transitions and phase transitions, and a more thorough understanding of the configuration space could shed light on these phenomena. Specifically, the distribution of critical points of the potential energy surface could constrain the differentiability of the configurational entropy, and therefore regulate the onset of a phase transition. In an effort to simplify the analysis, the base configuration space is often quotiented by various symmetries, e.g., rigid translations and permutations of particle labels. An approach to explicitly triangulate these quotient spaces is established in this work, using techniques from topological data analysis. Descriptors invariant to the desired symmetry groups are proposed, allowing the various quotient spaces to be parameterized. Two distance functions are provided, one induced by the quotient map and the other the Euclidean distance in the descriptor space. These allow the construction of explicit triangulations of the quotient spaces as $\alpha$-complexes, and thereby offer new approaches to studying the hard disk system. Specifically, the topological and geometric properties of the spaces can be directly evaluated as functions of disk radius. Some of the machinery developed is expected to be useful in other contexts as well, e.g., the proposed distance functions could be used to analyze the similarity of hard disk configurations generated by Monte Carlo simulations.

	The procedure to triangulate the configuration space is developed and applied to the simple but nontrivial cases of two hard disks in the square and hexagonal toruses. The first finding is that the use of a square or hexagonal torus does not substantially affect the topology of the quotient spaces except for the number of critical points of the tautological function $\tau$; the overall properties of the spaces are otherwise similar. The second finding is that the number and behavior of the critical points depends on the construction of the quotient space. For example, some of the index-$1$ critical points are identified with one another when the base configuration space is quotiented by rigid translations. The third finding is that the topology and the geometry of the quotient spaces change dramatically as additional symmetries are quotiented out. For example, the superlevel sets of $\tau$ can no longer be described as manifolds with boundaries, and instead need to be described as stratified spaces. The $\rho = 0$ configuration, which is not identified as a critical point in the context of classical Morse theory, consistently behaves as an index-$2$ critical point that closes the space.
	
	Even though this work considers only a pair of hard disks, extending and applying the techniques to the configuration spaces of more hard disks should be conceptually straightforward. The main obstacle is likely to be that the computational complexity of the distance defined in Eq.\ \ref{eq:unlabeled_distance} grows as $n!$ (the order of the permutation group). Another future direction could be to use the stratified Morse theory of Goresky and MacPherson \cite{goresky1988stratified} to more thoroughly analyze the effects of the quotients maps on the topology of the spaces.

	\begin{acknowledgments}
		O.B.E and J.K.M. were  supported  by  the  National  Science  Foundation under Grant No. 1839370.
	\end{acknowledgments}

	\appendix
	\section{Proof that $d_\Lambda$ is a metric}
	\label{sec:metric_proofs}
	
	This section proves that the distance function in Eq.\ \ref{eq:labeled_distance} is a metric. Let $\vec{x}$ be a column vector in $\RR^2$ and $\mat{P}$ be a projection matrix whose rows contain the unit vectors of the square or hexagonal domain as in Fig.\ \ref{fig:figure2}. By convention, the fundamental cell is defined as the set of points $\vec{x}$ such that $w_j \in [-0.5, 0.5)$ for all $j$ and $\vec{w} = \mat{P} \vec{x}$. Let $\config{p}, \config{q}, \config{r} \in \Lambda(n)$ be three distinct configurations in the following.
	
	First, the proposed distance function satisfies the identity of indiscernibles, or $d_\Lambda(\config{p},\config{q}) = 0 \iff \config{p} = \config{q}$.
	\begin{proof}
		That $\config{p} = \config{q} \implies d_\Lambda(\config{p},\config{q}) = 0$ is true by inspection. For the other direction, observe that $d_\Lambda(\config{p}, \config{q}) = 0$ implies that $\lVert \vec{p}_i - \vec{q}_i \rVert = 0$ for all $i$. Consider the $i$th disk, and drop the index in the following. 
		
		For the square torus in Fig.\ \ref{fig:figure2}, the geodesic distance reduces to
		\begin{align*}
		\lVert \vec{p} - \vec{q} \rVert &= \sqrt{a^2 + b^2} \\
		a &= \min{(\abs{p_x - q_x}, 1 - \abs{p_x - q_x})} \\
		b &= \min{(\abs{p_y - q_y}, 1 - \abs{p_y - q_y})}.
		\end{align*}
		That $\lVert \vec{p} - \vec{q} \rVert = 0$ implies that $a = 0$ and $b = 0$. Since $\vec{p}$ and $\vec{q}$ are assumed to be in the fundamental cell, $p_x, q_x \in [-0.5, 0.5)$ and $\abs{p_x - q_x} < 1$. Then $a = 0$ requires that $\abs{p_x - q_x} = 0$, or that $p_x = q_x$. $b = 0$ implies that $p_y = q_y$ by a similar argument, or that $\lVert \vec{p}_i - \vec{q}_i \rVert = 0$ if and only if $\vec{p}_i = \vec{q}_i$. Then $d_\Lambda(\config{p}, \config{q}) = 0 \implies \config{p} = \config{q}$.
		
		For the hexagonal torus in Fig.\ \ref{fig:figure2}, $\mat{P_h} = [1,0; 1/2, \sqrt{3}/2; -1/2, \sqrt{3}/2]$ and the geodesic distance reduces to
		\begin{align*}
			\lVert \vec{p} - \vec{q} \rVert &= \{\min[a^2 + b^2, (1 - a)^2 + b^2,\\
			&\qquad\qquad\! (a - 0.5)^2 + (b - \sqrt{3} / 2)^2] \}^{1 / 2}\\
			a &= \abs{p_x - q_x} \\
			b &= \abs{p_y - q_y}.
		\end{align*}
		Let $\mat{P}_h \vec{p} = \vec{t}$ and $\mat{P}_h \vec{q} = \vec{w}$. Since $\vec{p}$ and $\vec{q}$ are assumed to be in the fundamental cell, $t_j, w_j \in [-0.5, 0.5)$ for all $j$. The seven possible ways for $\lVert \vec{p} - \vec{q} \rVert = 0$ are shown in Table \ref{table:distance_hexagon}. Observe that only the first satisfies the assumption that $w_j \in [-0.5, 0.5)$ for all $j$, and therefore $\vec{t} = \vec{w}$. This implies that $\lVert \vec{p}_i - \vec{q}_i \rVert = 0$ if and only if $\vec{p}_i = \vec{q}_i$. Then $d_\Lambda(\config{p}, \config{q}) = 0 \implies \config{p} = \config{q}$.

		\begin{table}[h]
			\centering
			\begin{tabular}{||c | c | c | c | c | c||} 
				\hline
				$i$  & $q_x$ & $q_y$ & $w_1$ & $w_2$ & $w_3$	\\ [0.5ex] 
				\hline\hline
				1	& $p_x$			& $p_y$ 			& $t_1$ 	& $t_2$			& $t_3$  		\\
				2	& $p_x-1$		& $p_y$ 			& $t_1-1$ 	& $t_2-1/2$		& $t_3+1/2$	\\
				3	& $p_x+1$		& $p_y$ 			& $t_1+1$ 	& $t_2+1/2$		& $t_3-1/2$		\\
				4	& $p_x-1/2$		& $p_y-\sqrt{3}/2$ 	& $t_1-1/2$ & $t_2-1$		& $t_3-1/2$		\\
				5	& $p_x-1/2$		& $p_y+\sqrt{3}/2$ 	& $t_1-1/2$ & $t_2+1/2$		& $t_3+1$		\\
				6	& $p_x+1/2$		& $p_y-\sqrt{3}/2$ 	& $t_1+1/2$ & $t_2-1/2$		& $t_3-1$	\\
				7	& $p_x+1/2$		& $p_y+\sqrt{3}/2$ 	& $t_1+1/2$ & $t_2+1$		& $t_3+1/2$		\\[1ex]
				\hline
			\end{tabular}
			\caption{The seven possibilities that might have zero distance for the hexagonal torus. Recall that $t_1, t_2, t_3 \in (-0.5,0.5)$. Then, only the first one can be in the fundamental cell.}
			\label{table:distance_hexagon}
		\end{table}
		
	\end{proof}
	
	Second, the proposed distance function is symmetric, or $d_\Lambda(\config{p},\config{q}) = d_\Lambda(\config{q},\config{p})$, by the symmetry of $\lVert \vec{p}_i - \vec{q}_i \rVert$.
	
	Third, the proposed distance function satisfies the triangle inequality, or $d_\Lambda(\config{p}, \config{r}) \leq d_\Lambda(\config{p}, \config{q}) + d_\Lambda(\config{q}, \config{r})$.
	\begin{proof}
		The triangle inequality can be explicitly rewritten as $\sum_{i = 1}^n \lVert \vec{p}_i - \vec{r}_i \rVert \leq \sum_{i = 1}^n \lVert \vec{p}_i - \vec{q}_i \rVert + \sum_{i = 1}^n \lVert \vec{q}_i - \vec{r}_i \rVert$. Observe that the equation is true if the inequality holds separately for the $i$th term in each of the sums, and that this is true since $\lVert \cdot \rVert$ is the geodesic distance.
	\end{proof}

	\section{Relations between the $z_{\vec{k}}$}
	\label{sec:descriptor_proofs}
	
	The descriptors $z_{\vec{k}}$ in Eq.\ \ref{eq:z_k} are invariant under translation.
	\begin{proof}
		Let $\config{p}$ and $\config{q}$ be two configurations that differ by a translation $\vec{\Delta}$ such that $\vec{p}_i = \vec{q}_i + \vec{\Delta}$ for all $i$. Then
		\begin{align*} 
		z_{\vec{k}}(\config{p}) &= \bigg\lVert \sum_{j=1}^{n} e^{-2\pi i \vec{k} \cdot \vec{p}_j}\bigg\rVert \\ 
		&= \bigg\lVert \sum_{j=1}^{n} e^{-2\pi i \vec{k} \cdot (\vec{q}_j + \vec{\Delta})}\bigg\rVert \\
		&= \bigg\lVert e^{-2\pi i \vec{k} \cdot \vec{\Delta}} \sum_{j=1}^{n} e^{-2\pi i \vec{k} \cdot \vec{q}_j}\bigg\rVert \\
		&= \underbrace{\bigg\lVert e^{-2\pi i \vec{k} \cdot \vec{\Delta}} \bigg\rVert}_\text{1}
		\bigg\lVert \sum_{j=1}^{n} e^{-2\pi i \vec{k} \cdot \vec{q}_j}\bigg\rVert \\
		z_{\vec{k}}(\config{p}) &=  z_{\vec{k}}(\config{q})
		\end{align*}
	\end{proof}
	
	The descriptors $z_{\vec{k}}$ in Eq.\ \ref{eq:z_k} are invariant under inversion.
	\begin{proof}
		Let $\config{p}$ and $\config{q}$ be two configurations that differ by an inversion symmetry such that $\vec{p}_i = -\vec{q}_i$ for all $i$. Then
		\begin{align*} 
		c_{\vec{k}}(\config{p}) &= 
		\sum_{j = 1}^{n} e^{-2\pi i \vec{k} \cdot \vec{p}_j} = 
		\sum_{j = 1}^{n} e^{2\pi i \vec{k} \cdot \vec{q}_j} = 
		c_{\vec{k}}^\ast(\config{q}) \\ 
		z_{\vec{k}}(\config{p}) &= \sqrt{c_{\vec{k}}(\config{p}) c_{\vec{k}}^\ast(\config{p})} = \sqrt{c_{\vec{k}}^\ast(\config{q}) c_{\vec{k}}(\config{q})} = z_{\vec{k}}(\config{q})
		\end{align*}
	\end{proof}
	
	Finally, the descriptors $\hat{z}_{\vec{k}}$ in Eq.\ \ref{eq:z_hat_k} are not all independent. Let $\vec{x}$ and $\vec{a}$ respectively be the coordinates of a vector in the $xy$-coordinate system and $a_1a_2$-coordinate system. Let $\mat{T}$ be the forward transformation matrix such that $\vec{a} = \mat{T} \vec{x}$ and $\vec{x} = \mat{T}^{-1} \vec{a}$. 
		
		Let $\mathcal{L}$ be the set of symmetries of the tiling of the plane, and $\mat{L}$ be one of the corresponding matrices written using the $xy$-coordinate system. Let $\vec{x}$ be the position of a disk and $\vec{x}' = \mat{L} \vec{x}$ the position of the disk under the action of $L \in \mathcal{L}$. Then
		\begin{equation*} 
		\vec{a}' = \mat{T} \vec{x}' = \mat{T} \mat{L} \vec{x} = \mat{T} \mat{L} \mat{T}^{-1} \vec{a} = \mat{U} \vec{a}  
		\end{equation*}
		where $\mat{U} = \mat{T} \mat{L} \mat{T}^{-1}$. $\mat{U}$ is equivalent to $\mat{L}$, but is written using the $a_1a_2$-coordinate system.
		
		For the square torus, $\mat{T} = [1, 0; 0, 1]$ and $\mathcal{L}$ is the dihedral symmetry group of order eight ($D_4$). Equation \ref{eq:z_hat_k} is
		\begin{equation*}
		\hat{z}_{\vec{k}} = \frac{1}{8} \sum_{j = 1}^{8} z_{\vec{k}}^{L_j}
		\end{equation*}
		where $z_{\vec{k}}^{L_j}$ are the descriptors $z_{\vec{k}}$ of the configuration $L_j \config{x}$ with $L_j \in \mathcal{L}$. The $z_{\vec{k}}^{L_j}$ can instead be written as
		\begin{equation*} 
		z_{\vec{k}}^{L_j} = \bigg\lVert \sum_{l = 1}^{n} e^{-2\pi i 
			\vec{k} \cdot \mat{U}_j \vec{a}_l} \bigg\rVert = \bigg\lVert \sum_{l = 1}^{n} e^{-2\pi i 
			\vec{k}' \cdot \vec{a}_l} \bigg\rVert
		\end{equation*}
		where $\vec{k}' = \mat{U}_j^T \vec{k}$. Computing $\vec{k}'$ for the elements in $\mathcal{L}$ for the square torus yields
		\begin{align*}
		[p,q] &\sim [-p,q] \sim [q,-p] \sim [-q,-p ] \sim [-p,-q] \\
		&\sim [p,-q] \sim [-q,p] \sim [q,p].
		\end{align*}
		
		For the hexagonal torus, $\mat{T} = [1, -1/\sqrt{3}; 0, 2 / \sqrt{3}]$ and $\mathcal{L}$ is the dihedral symmetry group of order twelve ($D_6$). Repeating the procedure above yields
		\begin{align*}
			[p,q] &\sim [q,q-p] \sim [q-p,-p] \sim [-p,-q] \sim [-q,p-q] \\ 
			&\sim [p-q,p] \sim [-p,q-p] \sim [-q,-p] \sim [p-q,-q] \\
			&\sim [p,p-q] \sim [q,p] \sim [q-p,q].
		\end{align*}

\end{document}